# PERFORMANCE ANALYSIS OF BIO-INSPIRED ROUTING PROTOCOLS BASED ON RANDOM WAYPOINT MOBILITY MODEL

Vaibhav Godbole[*]

Department of Information Technology, Fr. Conceicao Rodrigues College of Engineering, India

[*]E-mail: vai.godbole@gmail.com

**ABSTRACT**

*A mobile ad hoc network (MANET) is a non-centralised, multihop, wireless network that lacks a common infrastructure and hence it needs self-organisation. The biggest challenge in MANETs is to find a path between communicating nodes, which is the MANET routing problem. Biology-inspired techniques such as ant colony optimisation (ACO) which have proven to be very adaptable in other problem domains, have been applied to the MANET routing problem as it forms a good fit to the problem. The general characteristics of these biological systems, which include their capability for self-organisation, self-healing and local decision making, make them suitable for routing in MANETs. In this paper, we discuss a few ACO based protocols, namely AntNet, hybrid ACO (AntHocNet), ACO based routing algorithm (ARA), imProved ant colony optimisation routing algorithm for mobile ad hoc NETworks (PACONET), ACO based on demand distance vector (Ant-AODV) and ACO based dynamic source routing (Ant-DSR), and determine their performance in terms of quality of service (QoS) parameters, such as end-to-end delay and packet delivery ratio, using Network Simulator 2 (NS2). We also compare them with well known protocols, ad hoc on demand distance vector (AODV) and dynamic source routing (DSR), based on the random waypoint mobility model. The simulation results show how this biology-inspired approach helps in improving QoS parameters.*

## 1. INTRODUCTION

Ad-hoc networks can be classified in three categories based on their applications; mobile ad-hoc networks (MANETs), wireless mesh networks (WMNs) and wireless sensor networks (WSN) (Mann and Mazhar, 2011). MANET is a collection of mobile nodes with a wireless network interface which forms a temporary network without the aid of any fixed infrastructure or centralised administration. Nodes within each other's transmission ranges can communicate directly, but nodes outside each other's range have to rely on other nodes to transmit the messages. Figure 1 shows a basic ad-hoc network. In this network, packet transmission from source to destination takes place without a base station.

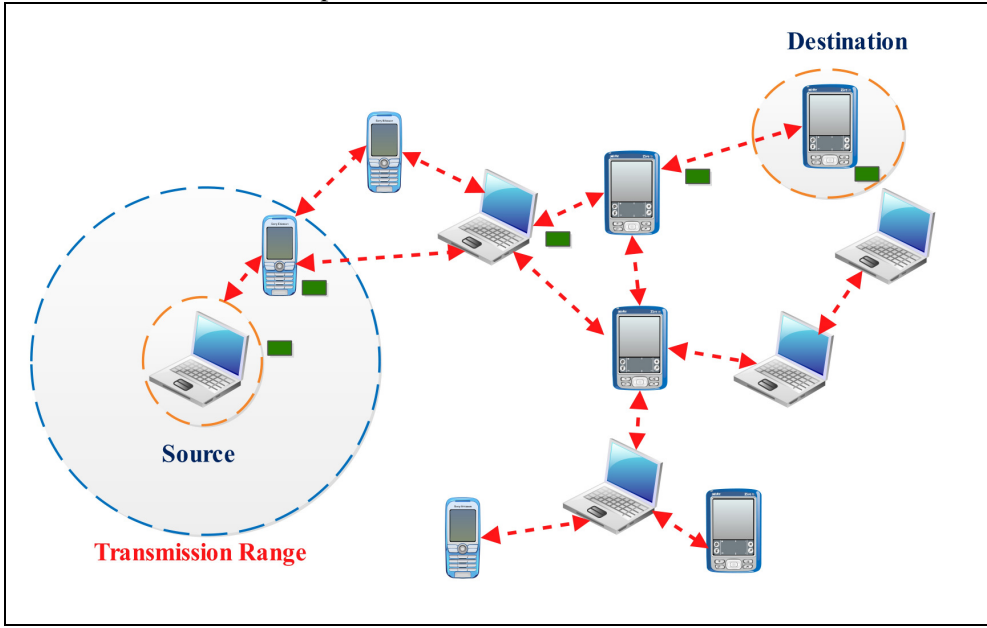

**Figure 1: A basic ad-hoc network.**
**(Source: Villalba *et al*. (2011))**

Research interest in MANETs has been growing in past few years, with the design of MANET routing protocols receiving significant attention. One of the reasons for this is that routing in MANETs is a particularly challenging task due to the fact that the topology of the network changes constantly, and paths which were initially efficient can quickly become inefficient or even infeasible. Moreover, control of information flow in the network is very restricted. This is because the bandwidth of the wireless medium is very limited, and as the medium is shared, nodes can only send or receive data if no other node is sending in their radio neighbourhood (Abolhasan *et al.*, 2004).

Basically, routing protocols in MANETs are classified into three categories; proactive, reactive and hybrid (Abolhasan *et al.*, 2004). Proactive routing protocols often need to exchange control packets among mobile nodes and continuously update their routing tables. Each node must maintain the state of the network in real time. This causes high overhead congestion of the network, which requires a lot of memory. The advantage of proactive protocols is that nodes have correct and updated information. Hence, when a path is required, it can be found directly in the memory and links can be established quickly. These protocols are intended to reduce broadcasting frequency while maintaining correct information for the routing table. Reactive routing protocols only seek a route to the destination when it is needed. The advantage of these protocols is that the routing tables located in the memory are not continuously updated. On the other hand, they have the disadvantage that they cannot establish connections in real time. The aim of these protocols is to save time in the route discovery process, since the reactive protocol is designed to reduce the latency which is critical in this kind of protocols. It also aims to avoid the maintenance of routes to prevent long delay (Singla & Kakkar, 2010).

It is therefore important to design protocols that are adaptive, robust and self-healing. Moreover, they should work in a localised way, due to the lack of central control or infrastructure in the network. Nature's self-organising systems, such as insect societies, termite hills, bee colonies, bird flocks and fish schools, provide precisely these features and hence have been a source of inspiration for the design of many routing algorithms for MANETs (Abdel-Moniem *et al.*, 2010). In this paper we discuss few ant colony optimisation (ACO) based routing protocols for MANETs, namely AntNet, hybrid ACO (AntHocNet), ACO based routing algorithm (ARA), improved ant colony optimisation routing algorithm for mobile ad hoc NETworks (PACONET), ACO based on demand distance vector (Ant-AODV) and ACO based dynamic source routing (Ant-DSR). We choose these protocols for analysis because all these protocols are based on forward ants (FANT) and backward ants (BANT) principle We compare these protocols based on quality of service (QoS) parameters such as end-to-end delay and packet delivery ratio, with conventional protocols such as ad hoc on-demand distance vector (AODV) and dynamic source routing (DSR), based on the random waypoint mobility model.

The random waypoint mobility model includes pause times between changes in direction and/or speed. A mobile node (MN) begins by staying in one location for a certain period of time (i.e., a pause time). Once this time expires, the MN chooses a random destination in the simulation area and a speed that is uniformly distributed between the minimum speed and maximum speed. The MN then travels toward the newly chosen destination at the selected speed. Upon arrival, the MN pauses for a specified time period before starting the process again (Camp & Davies, 2002). Figure 2 shows the travelling pattern of a MN using random waypoint mobility model.

## 2. ANT COLONY OPTIMISATION (ACO)

### 2.1 Similarities between Ad Hoc Networks and Ants

There are lots of similarities between ad hoc networks and ants, such as shown in Table 1. Ant based routing algorithms exhibit a number of desirable properties for ad hoc networks. The foraging behaviour of ants and bees, and the hill building behaviour of termites have inspired researchers in developing efficient routing algorithm for ad hoc networks (Gupta *et al.*, 2012).

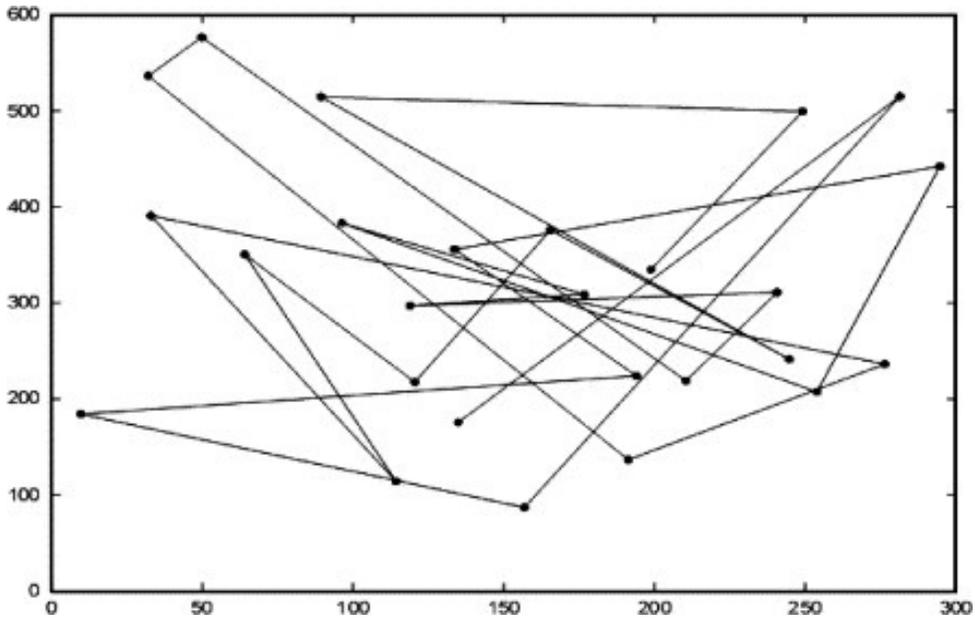

**Figure 2: Traveling pattern of a MN using the random way point mobility model. (Source: Camp & Davis, 2002)**

**Table 1: Comparison between ad hoc networks and ants. (Source: Gupta *et al.*, 2012)**

| Parameters | Ad-hoc networks | Ants |
|---|---|---|
| Physical structure | Unstructured, dynamic, distributed | Unstructured, dynamic, distributed |
| Origin of route | Route requests are sent from source to get local information | Pheromones are used to build new routes |
| Multipath support | Single path, partially multipath | Provide multipath |
| Basic system | Self-configuring, Self-organising | Self-configuring, Self-organising |

### 2.2 Ants in Nature

The main source of inspiration behind ACO is a behaviour that is displayed by certain species of ants in nature during foraging. It has been observed that ants are able to find the shortest path between their nest and a food source. The only way that this difficult task can be realised is through the cooperation between the individuals in the colony (Caro & Dorigo, 1998).

The key behind the colony level shortest path behaviour is the use of pheromone. This is a volatile chemical substance that is secreted by the ants in order to influence the behaviour other ants and of it. Pheromone is not only used by ants to find shortest paths, but is in general is an important tool that is used by many different species of ants (Caro & Dorigo, 1998).

Ants moving between their nest and a food source leave a trail of pheromone behind, and they also preferably go in the direction of high intensities of pheromone. We use the example situation depicted in Figure 3 to explain how this simple behaviour leads to the discovery of shortest paths. In our example, there are two possible paths between the ant nest and the food source, one of which is considerably shorter than the other. The first ants leaving the nest have no information available. They therefore choose their movements randomly. This leads to approximately 50% of the ants choosing the short path and 50% choosing the long path (Jha *et al*., 2011). All moving ants leave a trail of pheromone behind. The ants going over the short path reach the destination earlier than those going over the long path. Moreover, they can return faster. This leads temporarily to a higher pheromone concentration on the shortest path. Subsequent ants leaving the nest are attracted by this higher intensity, and go therefore preferably also over the shortest path. As this process continues, the majority of the ants eventually concentrate on the shortest path. However, it should be pointed out that the behaviour of the ants is never deterministic and hence, there will always remain a minority of ants that explore the longer path (Marwaha *et al.*, 2002).

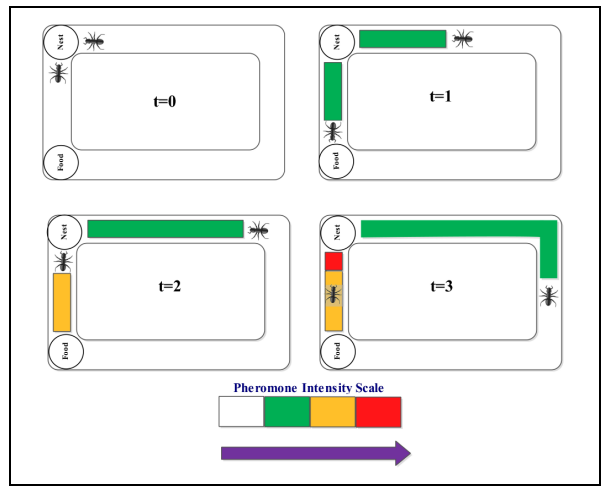

**Figure 3: The shortest path mechanism used by ants. The different colours indicate increasing levels of pheromone intensity. The scenario is depicted in successive time steps *t*.**
**(Source: Baluja & Davies, 1998)**

## 2.3 Routing in Ant Systems

The core of any network control system is routing which strongly affects the overall network performance. Routing deals with the problem of defining the path to forward incoming data traffic such that the overall network performance is maximised. At each node, data is forwarded according to the content of the routing table, maintains the information of source address, destination address, ant rip times etc. In this sense, a routing system can be seen as a distributed decision system (Jha *et al*., 2011). A variety of different classes of specific routing can be defined according to the different characteristics of processing, transmission components, traffic pattern and type of performance (Baluja & Davies 1998).

The following set of core properties of ACO characteristics for routing problems (Baluja & Davies 1998).

- provide traffic-adaptive and multipath routing
- rely on both passive and active information monitoring and gathering
- make use of stochastic components
- do not allow local estimates to have global impact
- set up paths in a less selfish way than in pure shortest path schemes favouring load balancing

These are all characteristics that directly result from the application of ACO's design guidelines, and in particular, from the use of controlled random experiments (the ants) that are repeatedly generated in order to actively gather useful non-local information about the characteristics of the solution set (i.e., the set of paths connecting all pairs of source-destination nodes in the routing case) (Baluja & Davies 1998).

## 3. ACO ALGORITHMS

In the nature, ants lay pheromone and so they produce pheromone trails between the nest and a food source. On a computer, the pheromone has been replaced by artificial stigmergy, the probabilities in the routing tables. To compute and update the probabilities, intelligent agents are introduced to replace the ants. There exist two kinds of agents, forward and backward agents. All forward and backward agents have the same structure. The agents move inside the network by hopping at every time step from a node to the next node along the existing links. The agents communicate with each other in an indirect way by concurrently reading and writing the routing tables on their way (Villalba & Orozco, 2010). In this section, we discuss based routing protocols for MANETs, namely AntNet, AntHocNet, ARA, PACONET, Ant-AODV and Ant-DSR.

### 3.1 AntNet

AntNet is a direct extension of the simple ACO algorithm (Di Caro, 1998). AntNet is even closer to the real ant's behaviour that inspired the development of the ACO meta-heuristic than the original ACO algorithms (Jha *et al*. 2011).

AntNet is conveniently described in terms of two sets of artificial ants, called as forward ant (FANT) and backward ant (BANT). Ants in each set possess the same structure, but they are situated differently in the environment; that is, they can sense different inputs and they can produce different, independent outputs. Ants communicate in an indirect way, according to the stigmergy paradigm, through the information they concurrently read and write on the network nodes they visit (Di Caro, 1998).

At regular intervals $\Delta t$ from every network node $s$, a FANT $F_{s \to d}$ is launched towards a destination node $d$ to discover a feasible, low-cost path to that node and to investigate the load status of the network along the path. FANTs share the same queues as data packets, so that they experience the same traffic load. Destinations are locally selected according to the data traffic patterns generated by the local workload: if $f_{sd}$ is a measure (in bits or in the number of packets) of the data flow $s \to d$, then the probability of creating at node $s$ a FANT with node $d$ as destination is given by (Villalba & Orozco, 2010):

$$P_{sd} = \frac{f_{sd}}{\sum_{i}^{n} f_{si}} \qquad (1)$$

The ant builds a path using the following steps:

1. At each node $i$, each FANT headed toward a destination {\it d} selects the node $j$ to move to, choosing among the neighbours it did not already visit, or over all the neighbours in case all of them had previously been visited. The neighbour $j$ is selected with a probability $P_{ijd}$ computed as the normalised sum of the pheromone $\tau_{ijd}$ with a heuristic value $n_{ij}$ taking into account the state (the length) of the $j^{th}$ rank link queue of the current node $i$:

$$P_{ijd} = \frac{\tau_{ijd} + \alpha n_{ij}}{1 + \alpha(|N_i| - 1)} \qquad (2)$$

   The heuristic value $n_{ij}$ is a [0, 1] normalised values function of the length $q_{ij}$ (in bits waiting to be sent) of the queue on the link connecting the node $i$ with its neighbour $j$:

$$n_{ij} = 1 - \frac{q_{ij}}{\sum_{l}^{|N_i|} q_{il}} \qquad (3)$$

The value of $α$ weighs the importance of the heuristic value with respect to the pheromone values stored in the pheromone matrix $T$. The value $h_{ij}$ reflects the instantaneous state of the node's queues and, assuming that the queue's consuming process is almost stationary or slowly varying, $h_{ij}$ gives a quantitative measure associated with the queue waiting time (Caro & Dorigo, 1998).

2. When the destination node $d$ is reached, the agent $F_{s \rightarrow d}$ generates another agent, BANT $B_{d \rightarrow s}$, transfers to all of its memory, and is deleted. A FANT is also deleted of its lifetime and becomes greater than a value *max_life* before it reaches its destination node, where *max_life* is a parameter of the algorithm.

3. The BANT takes the same path as that of its corresponding FANT, but in the opposite direction. BANTs do not share the same link queues as data packets; they use higher-priority queues reserved for routing packets, because their task is to quickly propagate to the pheromone matrices the information accumulated by the FANTs (Sujatha & Harigovindan, 2010).

4. The re-enforcement factor $r$ is defined as the ratio of travel time of an ant at a specific node to the travel time of all ants at that node. The value of $r$ is such that $0<r<1$. This factor is pre-defined in the AntNet algorithm.

## 3.2 Hybrid ACO (AntHocNet)

AntHocNet (Di Caro *et al*., 2005) combines the typical path sampling behaviour of ACO algorithms with a pheromone bootstrapping mechanism. AntHocNet is a hybrid algorithm. It is reactive in the sense that a node only starts gathering routing information for a specific destination when a local traffic session needs to communicate with the destination and no routing information is available. It is proactive because as soon as the communication starts, and for the entire duration of the communication, the nodes proactively keep the routing information related to the on-going flow up-to-date with network changes for both topology and traffic. The algorithm tries to find paths characterised by minimal number of hops, low congestion and good signal quality between adjacent nodes.

Nodes in AntHocNet forward data stochastically. When a node has multiple next hops for the destination d of the data, it randomly selects one of them with probability $P_{nd}$. $P\$_{nd}$, which is calculated as follows:

$$P_{nd} = \frac{(T_{nd}^i)^\beta}{\sum_{j \in N_d^n} (T_{jd}^i)^\beta} \quad \beta >> 1 \tag{4}$$

where $N^i_d$ is the set of neighbours of *i* over which a path to *d* is known, and *β* is a parameter value which can control the exploratory behaviour of the ants.

The probabilistic routing strategy leads to data load spreading according to the estimated quality of the paths. When a path is clearly worse than others, it will be avoided, and its congestion will be relieved. Other paths will get more traffic, leading to higher congestion, which will make their end-to-end delay increase. By continuously adapting the data traffic, the nodes try to spread (Lin & Shao, 2010).

A node which receives multiple copies of the same ant only accepts the first and discards the others. When a FANT arrives at destination, it goes backward, updates the pheromone tables at the nodes, indicating a path between s and d, and triggers the sending of data packets from the traffic session. In this way, only one path is set up initially. During the course of the communication session, additional paths are added and / or removed via a proactive path maintenance and exploration mechanism. This is implemented through a combination of ant path sampling and slow-rate pheromone diffusion and bootstrapping, which mimics pheromone diffusion in nature. This way, promising pheromone is checked out, and if the associated path is there and has the expected good quality, it can be turned into a regular path available for data (Di Caro *et al.*, 2005).

### 3.3    ACO Based Routing Algorithm (ARA)

ARA (Gunes & Sorges, 2002) is a purely reactive MANET routing algorithm. It does not use any HELLO packets to explicitly find its neighbours. . HELLO packets are sent by the routers to compute the time delay to send and receive datagrams to and from its neighbors. A HELLO packet also consists of clock and timestamp information. When a packet arrives at a node, the node checks it to see if routing information is available for destination *d* in its routing table. From Figure 4, we can see that route discovery is done either by the FANT flood technique or FANT forward technique. In the FANT flooding scheme, when a FANT arrives to any intermediate node, the FANT is flooded to all its neighbours. If found, it forwards the packet over that node; if not, it broadcasts a FANT to find a path to the destination. By introducing a maximum hop count on the FANT, flooding can be reduced. In the FANT forwarding scheme, when a FANT reaches an intermediate node, the node checks its routing table to see whether it has a route to the destination over any of its neighbours. If such a neighbour is found, the FANT is forwarded to only that neighbour; else, it is flooded to all its neighbours as in the flood scheme. In ARA, a route is indicated by a positive pheromone value in the node's pheromone table over any of its neighbours to the FANT destination. When the ant reaches the destination it is sent back along the path it came, as a backward ant as shown in Figure 4(b). All the ants that reach the destination are sent back along their path. Nodes modify their routing table information when a backward ant is seen according to number of hops the ant has taken. When a route is found, the packet is forwarded over the next hop stochastically.

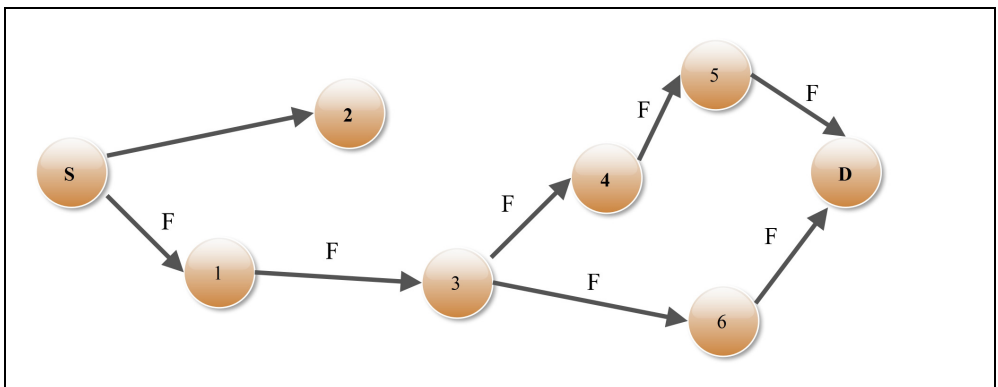

(a)

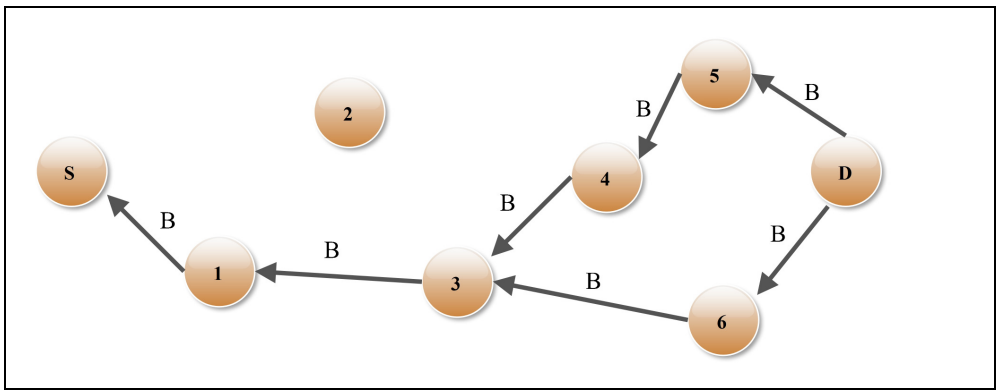

(b)

**Figure 4: Route discovery phase: (a) A FANT *F* is send from the sender *S* toward the destination node *D*. The forward ant is relayed by other nodes, which initialise their routing table and the pheromone values. (b) The BANT *B* has the same task as *F*. It is sent by *D* towards *S*.**
**(Source: Gunes *et al.*, 2002)**

### 3.4 imProved ant colony optimisation routing algorithm for mobile ad hoc NETworks (PACONET)

PACONET (Osagie & Thulasiram, 2008) is a routing protocol for mobile ad-hoc networks inspired by the foraging behaviour of ants. It uses the principles of ACO routing to develop a suitable problem solution.

The PACONET protocol is based on the parallel ACO algorithm. The availability of parallel architectures at low cost has widened the interest for the parallelisation of algorithms and metaheuristics. When developing parallel genetic algorithms and parallel ACO algorithms, it is common to adopt the strategy of information exchange that plays a major role in the algorithms. Solutions, pheromone matrices, and parameters have been tested as the object of such an exchange (Gunes & Sorges, 2002).

The fundamental principle of parallel ant colony algorithm (Middendorf *et al.*, 2002; Chengyong Liu & Xiang, 2008) is to divide $M$ ants into $P$ ant colonies. Normally, the numbers of every ant colony are the same, viz. $N_a = M/P$. In the algorithm development, each colony is distributed to a processor, and then, the ant colony can search the best solution independently. In order to avoid local optimisation in some processor when the ant colony is doing the job, the processors should carry out the information exchange each other in the fixed condition (i.e. time interval, etc.) The FANT explores the paths of the network in a restricted broadcast manner in search of routes from a source to a destination. The BANT establishes the path information acquired by the FANT.

When a source node $S$ wishes to communicate with a destination node $D$ for which it has no route information, it sends out a FANT to all its neighbours in search of the destination node. When a FANT from $S$ traveling to $D$, arrives at a node $v$, the FANT determines its path or next hop neighbour by looking at the node's routing table. It considers the node's neighbours by looking at the rows against the columns in the routing table to select the best path from a neighbouring node to $D$ rather than the best link between itself and its neighbour.

The FANT will consider the pheromone concentration only when all neighbours in column have been visited. The purpose of this is to ensure that all possible paths are explored to find the best path towards the destination. The node with the highest pheromone is chosen as the next hop after the FANT has determined that it has not visited the node before. This is to avoid the ant travelling in cycles. The FANT maintains a list of all nodes visited on its journey to $D$ for this purpose. The FANT keeps in memory the total time $T$ it has travelled. When a next hop node $v_j$ is selected from $v_i$ the FANT moves to $v_j$ and updates the pheromone entry for $(v_i, S)$ in $v_j$'s routing table using the following equation:

$$\delta(v_j, v_s) = \delta(v_i, v_j) + \frac{\varepsilon}{T(v_s, v_i) + w(v_i, v_j)} \qquad (5)$$

where $\varepsilon$ is a user defined run time parameter; $\delta(v_i, v_j)$ and $w(v_i, v_j)$ represent the pheromone value on each edge and time period respectively for which the links are in connection. For all the other nodes in the source column, the pheromone values are decremented by the following equation:

$$\delta(v_j, v_s) = (1 - \xi)\delta(v_j, v_s), \forall l \neq i \qquad (6)$$

where $\zeta$ is evaporation rate of the pheromone, which is also determined by the user.

The total time of the path just traversed is recorded as $T(v_s, v_i) + w(v_i, v_j)\}$. When the FANT reaches the destination, a corresponding BANT is created with the source of the FANT as its destination. The BANT travels towards its destination using the list of visited nodes acquired from the FANT while updating the pheromone concentration for the destination column. That is, to update an entry ($v_b$,$v_D$) for an ant at node $v_k$, travelling backwards from $v_b$ we look at the rows of $v_b$'s neighbouring nodes and column:

$$\delta(v_b, v_D) = \delta(v_b, v_D) + \frac{\varepsilon}{T'} \qquad (7)$$

where $T'$ is $T(v_s,v_d) - T(v_s,v_k)$. The advantage of performing this update is that it makes it easy to determine the best available path reachable from a source and to find a path easily when another ant considers the source as its destination.

### 3.5 ACO Based on Demand Distance Vector (Ant-AODV)

Ant-AODV (Marwaha *et al*., 2002; Abdel-Moniem & Hedar, 2010) is a hybrid protocol that is able to provide reduced end-to-end delay and high connectivity as compared to AODV. AODV does the reactive part and an ant-based approach does the proactive one. The main goal of the ant algorithm here is to continuously create routes in the attempt to reduce the end-to-end delay and the network latency, increasing the probability of finding routes more quickly, when required. Ant-AODV's artificial pheromone model is based on the number of hops and its goal is to discover the network topology, without any other specific functions, as opposed to most ACO algorithms. Route establishment in conventional ant-based routing techniques is dependent on the ants visiting the node and providing it with routes. The nodes also have capability of launching on-demand route discovery to find routes to destinations. The use of ants with AODV increases the node connectivity (the number of destinations for which a node has unexpired routes), which in turn reduces the amount of route discoveries and also the route discovery latency. This makes the Ant-AODV hybrid routing protocol suitable for real-time data and multimedia communications. Ant-AODV uses route error (RERR) messages to inform upstream nodes of a local link failure similar to AODV. The routing table in Ant-AODV is common to both the ants and AODV. Frequent HELLO broadcasts are used to maintain a neighbour table.

### 3.6 ACO Based Dynamic Source Routing (Ant-DSR)

Ant-DSR (Fenouche & Mellouk, 2007) is a reactive protocol that implements a proactive route optimisation method through constant verification of cached routes. This approach increases the probability of a given cached route expressed through the network reality. Mobile nodes are required to maintain route caches that contain the source routes of which the mobile node is aware. Entries in the route cache are continuously updated as new routes are learnt. The protocol consists of two major phases; route discovery and route maintenance. In Ant-DSR, the FANT and BANT packets are added in the route request and reply of DSR respectively. FANTs are used to explore new paths in the network, and measure the current network state for instance by trip times, hop count or Euclidean distance travelled. BANTs serve the purpose of informing the originating node about information collected by the FANTs.

## 4. PERFORMANCE EVALUATION

### 4.1 Simulation Parameters

Network Simulator 2 (NS2) is an open-source network simulator, which is used by the researchers to analyse performance of wired and wireless networks. Due to open source nature of NS2, new protocols can be implemented by any individual. Once simulation is done, a trace file is generated. We have used awk scripts to analyse the trace files to obtain QoS parameters such as end to end delay and packet delivery ratio. Table 2 shows the simulation parameters for the ACO protocols used in this study.

Table 2: Simulation parameters for the ACO protocols used in this study.

| Protocol | Parameters | | | | | |
|---|---|---|---|---|---|---|
| | No. of Nodes | Network Area ($m^2$) | Simulation Time (s) | The Radio Propagation Range (m) | Packet Size | $r$ |
| AntNet | 50 | 500 X 500 | 300 | 300 | 27 bytes. | 0.1 |
| AntHocNet | 50 | 500 X 500 | 300 | 300 | 27 bytes. | - |
| ARA | 50 | 500 X 500 | 300 | 300 | 1 packet of 64 KB /s | - |
| PACONET | 100 | 500 X 500 | 120 | 300 | 4 packets of 64 KB /s | - |
| Ant-AODV | 100 | 500 X 500 | 600 | 300 | 6 packets of 1000 bytes /s | - |
| Ant-DSR | 100 | 500 X 500 | 600 | 300 | 6 packets of 1000 bytes /s | - |

### 4.2 Discussion

Every trace file generated by NS2 during the simulation of AntNet is of very large size (around 300 MB). Hence, due to this practical difficulty, we have fixed packet size of 27 bytes, so that the trace file size will not increase too much. From Figure 5(a), we can see that the average end-to-end delay for AntNet is better (lower) than AODV, for varying node velocities. From Figure 6(a) we can see that the packet delivery ratio for AntNet is almost equal to 1, whereas for AODV, the packet delivery ratio is less than 1, for varying node velocities. Hence, AntNet performs better than AODV in terms of end-to-end delay and packet delivery ratio.

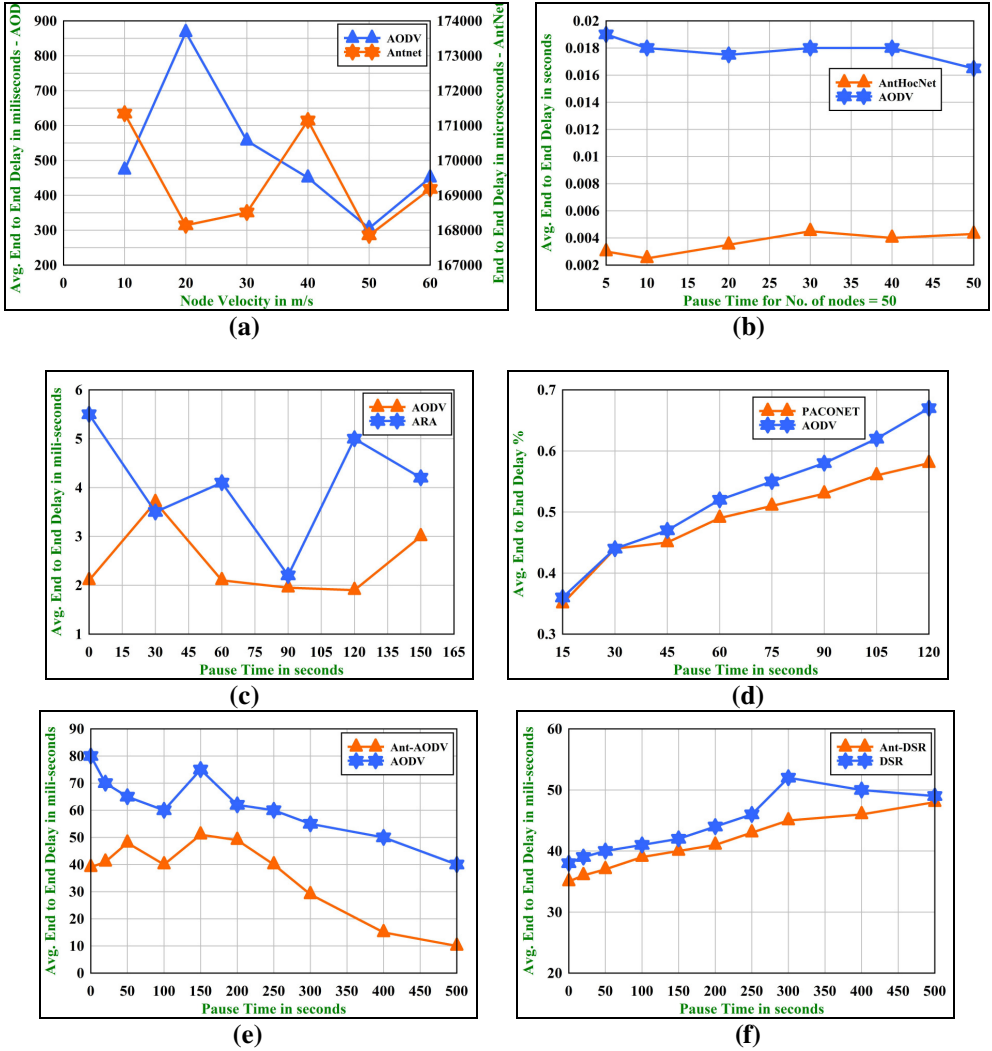

Figure 5: Average end-to-end delay for: (a) AntNet (b) AntHocNet (c) ARA (d) PACONET (e) Ant-AODV (f) Ant-DSR.

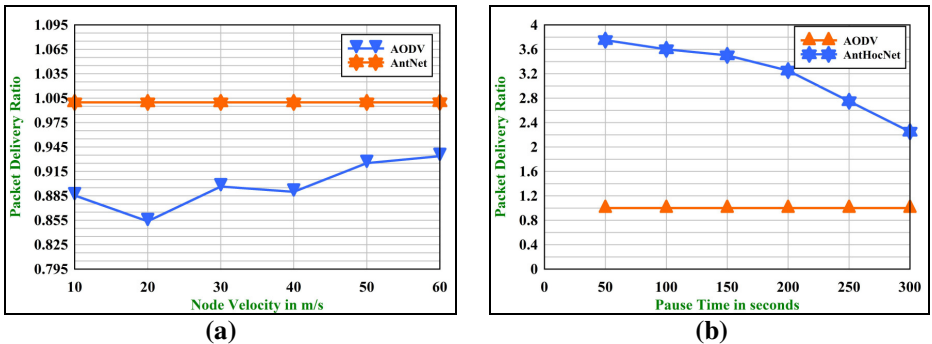

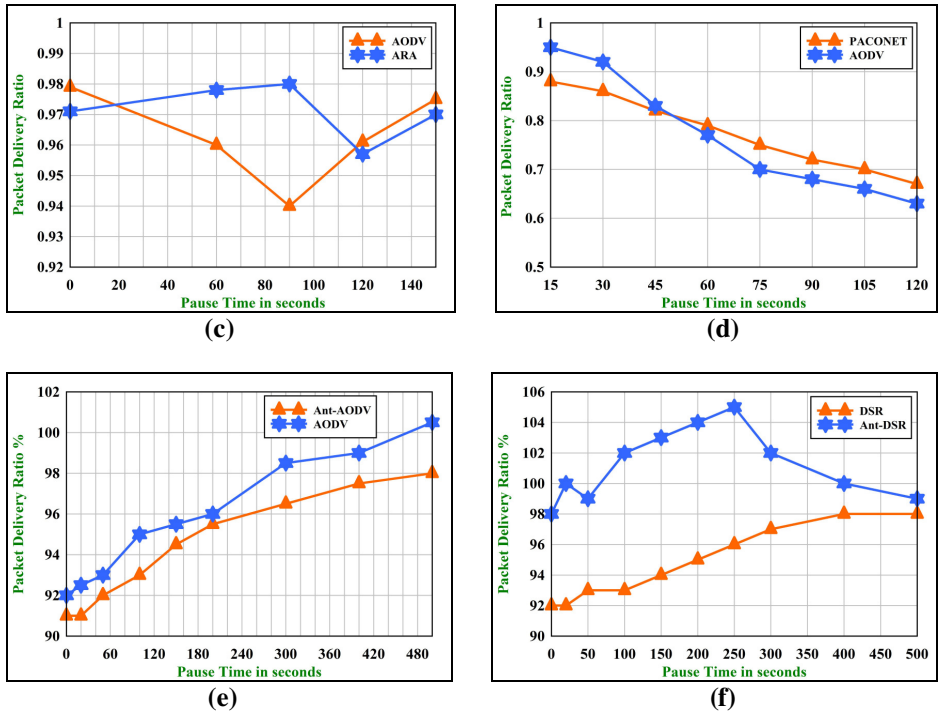

**Figure 6: Packet delivery ratio for: (a) AntNet (b) AntHocNet (c) ARA (d) PACONET (e) Ant-AODV (f) Ant-DSR.**

From Figure 5(b), we can see that AntHocNet has lower average end-to-end delay as compared to AODV. It is observed that for higher pause times, the end-to-end delay increases. This is due to the fact that under the Random Waypoint model, the nodes are concentrated more in the centre of the network rather than at the edges, especially for lower pause times. From Fgure 6(b), it can be seen that AntHocNet outperforms AODV in terms of packet delivery ratio. This is because AntHocNet uses a lot of different kinds of ant packets, such as FANT, BANT, etc., in order to adapt to changing MANET environments and form many optimal routes, thereby reducing the number of packet drops.

From Figure 5(c), we can see that the end-to-end to end delay for AODV is smaller than ARA. This is due to the fact that while AODV uses no route maintenance mechanism other than a timeout to delete stale routes, ARA uses a route maintenance mechanism to gradually modify the "freshness" of the routes. In addition, the route selection exponent makes the ant route selection equations more sensitive to changes in pheromone values. This change in pheromone values is indirectly indicative of the topology of the MANET and causes the ant route selection equations to select varied routes. Hence, from Figure 6(c), we can see that the packet delivery ratio of ARA is better than that of AODV. For lower mobility, AODV performs better than ARA.

In Figure 5(d), where the node pause time is varied, AODV shows higher overall delay. The delay in PACONET is usually significant at the start of the simulation because of the initial search for routes. Intermediate nodes in AODV are able to respond to route requests, thus saving time for path discovery. In PACONET, the source node has to wait until it gets a BANT sent by the destination. The initial peak in end-to-end delay is what causes AODV to outperform PACONET at higher node speeds. Furthermore, unlike most protocols, PACONET does no broadcasting which limits the extent of path discovery. In Figure 6(d), where the packet delivery ratio is measured at varying node pause times, both algorithms show improved performance. Their performance actually alternates; AODV starts out better but PACONET finishes with higher delivery ratio for longer pause times.

It is evident from the simulation results shown in Figures 5(e) and 6(e) that by combining ant-like mobile agents with the on-demand route discovery mechanism of AODV, the Ant-AODV hybrid routing protocol would give reduced end-to-end delay and high packet delivery ratio at large pause times. The high packet delivery fraction in Ant-AODV is because it makes use of link failure detection and route error messages. Whereas in the case of ant-based routing, there is no such feature and hence, the source nodes keeps on sending packets unaware of the link failures. This leads to a large amount of data packets being dropped, which reduces the packet delivery fraction and the throughput. It can be observed that end-to-end delay is considerably reduced in Ant-AODV as compared to AODV. As ants help in maintaining high connectivity in Ant-AODV, the packets need not wait in the send buffer until the routes are discovered. Even if the source node does not have a ready route to the destination, due to the increased connectivity at all the nodes, the probability of it receiving replies quickly from nearby nodes is high, resulting in reduced route discovery latency. Ant-AODV is able to provide reduced end-to-end delay and high connectivity as compared to AODV. As a result of increased connectivity, the number of route discoveries is reduced and also the route discovery latency. This makes Ant-AODV suitable for real-time data and multimedia communications. Table 3 summarises the differences between AODV and Ant-AODV.

**Table 3: Comparison between AODV and Ant-AODV.**

| Parameters | AODV | Ant-AODV |
|---|---|---|
| Routing type | Purely reactive | Hybrid |
| End-to-end delay | High | Low |
| Connectivity | Low | High |
| Route type | Single path | Multipath |
| Overhead | Low | High |

From Figure 5(f), we can see that the average end-to-end delay is reduced with Ant-DSR, while from Figure 5(f), we can see that the packet delivery ratio for Ant-DSR is higher than that of DSR. This is mainly due to the addition of the delay pheromone in the RREQ and RREP packets. The reduction in delay is at its maximum (15-20 %) when the pause time reaches beyond 300 s. Both protocols have the same delay for higher pause times. The packet delivery ratio shows an improvement over DSR, where it is high for low pause time. It can be seen that an increase in node speed results in significant decrease in both the protocols due to more link breakages. We can say that Ant-DSR produced better results than DSR in terms of packet delivery ratio and end-to-end delay. Table 4 summarises the differences between AODV and Ant-AODV.

Table 4: Comparison between DSR and Ant-DSR.

| Parameters | DSR | Ant-DSR |
|---|---|---|
| Routing type | Reactive | Reactive |
| End-to-end delay | High | Low |
| Energy and jitter | Low | High |
| Through | Low | High |
| Overhead | Low | High |

### 4.3 Summary of ACO Based Protocols for MANETs

Table 5 summarises the ACO based protocols used in this study in terms of routing and path types. Proactive routing protocols maintain routes to all destinations, regardless of whether or not these routes are needed. In order to maintain correct route information, a node must periodically send control messages. Therefore, proactive routing protocols may waste bandwidth. The main advantage of this category of protocols is that hosts can quickly obtain route information and establish a session. Reactive routing protocols can dramatically reduce routing overhead because they do not need to search for and maintain the routes on which there is no data traffic (Singla & Kakkar, 2010).

Table 5: Summary of ACO based protocols used in this study.

| Protocol | Routing type | Path type |
|---|---|---|
| AntNet | Proactive | Single |
| AntHocNet | Hybrid | Single |
| ARA | Reactive | Multipath |
| PACONET | Reactive | Single |
| Ant-AODV | Hybrid | Multipath |
| Ant-DSR | Reactive | Broadcast |

Hybrid methods combine proactive and reactive methods to find efficient routes, without much control overhead. In general, hybrid routing's flexibility allows the network operator to adjust the protocol operation to match the network's current mission and state. For example, a purely proactive operation might be used in relatively static networks, such as inter-ship links. In contrast, purely reactive routing might be used in dynamic networks such as clouds of tactical unmanned aerial vehicles (UAVs), and networks of ground-based sensors that have strict low probability of detection (LPD) requirements. These protocol adjustments could occur without changing the network software or "rebooting" any of the underlying MANET routers (Sholander *et al.*, 2002).

Single path routing is based on single route establishment between sourceand destination. In this routing, the packet is transmitted to the destination using a single route (Abolhasan *et al*., 2004). Multipath routing gives the choice to the source to choose the path between various available paths between source and destination by taking advantage of the connectivity redundancy of the underlined network (Di Caro, 1998). Broadcast routing is when a single device is transmitting a message to all other devices in a given address range. This broadcast could reach all hosts on the subnet, all subnets, or all hosts on all subnets. Broadcast packets have the host (and/or subnet) portion of the address set to all ones (Abolhasan & Dutkiewicza, 2004).

## 5. CONCLUSION AND FUTURE WORK

In this paper, we reviewed a few ACO based protocols, namely Antnet, AntHocNet, ARA, PACONET, Ant-AODV and Ant-DSR. We obtained QoS parameters such as end-to-end delay and packet delivery ratio for these protocols. We also compared these protocols with conventional protocols, AODV and DSR, based on the random waypoint mobility model. Our results shows that ACO based protocols perform better than conventional protocols in terms of end-to-end delay and packet delivery ratio.

Our research findings may be useful for researchers who wish to modify the existing ACO based protocols. The results obtained in this research can be used for comparison with the modified protocols.
For future work, the performance of these protocols can be analysed using other mobility models such as pursue, random direction and nomadic community mobility models. More critical performance evaluations of these protocols shall be done on the basis of simulations and other performance metrics such as routing overhead, route cost and normalised routing load.

## ACKNOWLEDGEMENT

The authors are grateful to the management of Fr. Conceicao Rodrigues College of Engineering, Fr. Agnel Technical Education Complex, Bandra (W), Mumbai: 400050, for allowing to use their infrastructure for experimentations.

## REFERENCES

Defence S & T Technical Bulletin, Science & Research Technology Institute for Defence (STRIDE), Vol. 5, No. 2, November 2012, pp. 114-134, ISSN: 1985-6571Abdel-Moniem, A.M. & Hedar, A. (2010). An ant colony optimizsation algorithm for the mobile ad hoc network routing problem based on AODV protocol . *Tenth Intl. Conf. Int. Syst. Design App.*, pp. 1332-1337.

Gupta, A., Sadawarti, H. & Verna, A. (2012). MANET routing protocols based on ant colony optimisation. *Intl. J. Modeling & Optimisation.*,2(1): pp. 42-49.

Abolhasan, M., Wysocki, T. & Dutkiewicz, E. (2004). A review of routing protocols for mobile ad hoc networks. *J. Ad Hoc Networks*, **2**: 1-22.

Aissani, M,. Fenouche, M. & Sadour , F. (2007). Ant-DSR: Cache maintenance based routing protocol for mobile ad-hoc networks. *Third Adv. Int. Conf. Telecom.,* pp. 35-40.

Baluja, S., & Davies, S. (1998). Fast probabilistic modeling for combinatorial optimizsation. *Fifteenth National Conf. AI,* pp. 469-476.

Camp, T., Boleng, J. & Davies, V. (2002). A survey of mobility models for ad hoc network research. *Wirel. Commun. Mobile Comput.*, **2**: 483-502.

Caro, G.D. & Dorigo, M. (1998). Mobile agents for adaptive routing. *Thirty-First Hawaii Intl. Conf. Syst. Sci.*, pp. 74-83.

Chengyong Liu, L.L. & Xiang, Y. (2008). Research of multi-path routing protocol based on parallel ant colony algorithm optimizsation in mobile ad hoc networks. *Fifth IEEE Intl. Conf. Inf. Tech.*, pp. 1006-1010.

Di Caro, G.A. (1998). Two ant colony algorithms for best-effort routing in datagram networks. *Tenth Intl. Conf. Parallel Distrib. Comput. Syts.*, pp. 28-31.

Di Caro, G.A., Ducatelle, F. & Gambardella, L.M. (2005). AntHocNet: An adaptive nature-inspired algorithm for routing in mobile ad hoc networks. *Eur. T. Telecommun.* , **16**: 443–455.

Gunes, M. & Sorges, U. (2002). ARA-the ant colony based routing algorithm for MANETs. *Intl. Conf. Parallel Proc. Workshops*, pp. 79-85.

Gupta, A.K., & Sadawarty, H. (2010). Performance analysis of AODV, DSR and TORA routing protocols. *J. Eng. Tech.,* **2**: 226-221.

Jha, K. Khetarpal, M. & Sharma, M. (2011). A survey of nature inspired routing algorithms for MANETs. *Third IEEE Intl. Conf. Electron. Comp. Tech.*, pp. 16-24.

Lin, N., & Shao, Z. (2010). Improved Ant Colony Algorithm for Multipath Routing Algorithm Research. *Intl. Symposium Intel. Inf. Process. Trusted Comput.*, pp. 651-655.

Mann, F. & Mazhar, M. (2011). MANET routing protocols vs mobility models: A performance evaluation.. *Third Intl. Conf. Ubiq. Fut. Networks*, pp. 179-184.

Middendorf, M., Reischle, F. & Schmeck, H. (2002). Multi colony ant algorithms. *J. Heuristics,* **8***:* 305-320.

Marwaha, S., Tham, C.K. & Srinivasan, D. (2002). A novel routing protocol using mobile agents and reactive route discovery for ad hoc wireless networks. *10th IEEE Intl. Conf. Networks*, pp. 311-316.

Marwaha, S., Tham, C.K. & Srinivasan, D. (2002). Mobile agents based routing protocol for mobile ad hoc networks. *Global Telecomm. Conf.,* pp. 163-167.

Osagie, E., & Thulasiram, R. (2008). PACONET: imProved ant colony optimizsation routing algorithm for mobile ad hoc NETworks. *22$^{nd}$ Intl. Conf. Adv. Inf. Networking App.*, pp. 204-211.

Perkins, C., Belding-Royer, E. & Das, S. (2003). Ad hoc on-demand distance vector (AODV) routing. RFC 3561. Available online at: http://www.ietf.org/rfc/rfc3561.txt (Last accessed date: 20 September 2012).

Singla, V. & Kakkar, P. (2010). Traffic Ppattern based performance comparison of reactive and proactive protocols of mobile ad-hoc networks, *J. Comp. App. ,* **5**:16-20.

Sujatha, B.R.& Harigovindan, V.P. (2010). The research and improvement of AntNet algorithm. *Second Intl. Asia Conf. Informatics Control, Autom. Robotics*, pp. 505-508.

Sholander, P., Yankopolus A., Coccoli, P. & Tabrizi, S.S. (2002). Experimental comparison of hybrid and proactive routing protocols for MANETs. *MILCOM 2002*, pp. 513-518.

Villalba, L.J.G., Matesanz, J.G., Orozco, A.L.S. & Díaz, J.D.M. (2011). Auto-configuration protocols in mobile ad hoc networks. *Sensors*, **11**: 3562-3666.

Villalba, L.J.G. & Orozco, A.L. (2010). Bio-inspired routing protocol for mobile ad hoc networks. *IET Comm.,* **4**: 2187-2195.